\documentclass[superscriptaddress,amsmath,amssymb]{revtex4-2}

\usepackage{diagbox}
\usepackage{color}
\usepackage[table]{xcolor}
\usepackage{multirow}
\usepackage{tabularx}
\usepackage{graphicx}
\usepackage{bm}
\usepackage{lettrine}
\usepackage{lmodern}
\begin{document}

\title{Missed ferroelectricity in methylammonium lead iodide}

\author{Wen-Yi Tong}
\affiliation{Theoretical Materials Physics, Q-MAT, CESAM, Universit\'e de Li\`ege, B-4000 Li\`ege, Belgium}
\author{Jin-Zhu Zhao}
 \email{zhaojz@m.scnu.edu.cn}
\affiliation{Guangdong Provincial Key Laboratory of Quantum Engineering and Quantum Materials,Guangdong-Hong Kong Joint Laboratory of Quantum Matter,  School of Physics and Telecommunication Engineering,South China Normal University, Guangzhou 510006, P. R. China}
\affiliation{Center for Computational Science and Engineering, Southern University of Science and Technology, Shenzhen 518055, P. R. China}
\author{Philippe Ghosez}
 \email{Philippe.Ghosez@uliege.be}
\affiliation{Theoretical Materials Physics, Q-MAT, CESAM, Universit\'e de Li\`ege, B-4000 Li\`ege, Belgium}

\begin{abstract}
\textbf{
Methylammonium lead iodide, as related organometal halide perovskites, emerged recently as a particularly attractive material for photovoltaic applications. The origin of its appealing properties is sometimes assigned to its potential ferroelectric character, which remains however a topic of intense debate. Here, we rationalize from first-principles calculations how the spatial arrangement of methylammonium polar molecules is progressively constrained by the subtle interplay  between their tendency to bond with the inorganic framework and the appearance of iodine octahedra rotations inherent to the perovskite structure. The disordered tetragonal phase observed at room temperature is paraelectric. We show that it should a priori become ferroelectric but that iodine octahedra rotations drive the system toward an antipolar orthorhombic ground state, making it a missed ferroelectric.
}
\end{abstract}

\maketitle

\lettrine[lines=3,loversize=0.1,nindent=-0.8pt]{\textbf{O}}{}rganometal halide perovskites (OMHPs) of chemical formula ABX$_{3}$ -- with A an organic molecule, B a metal atom and X a halogen atom -- possess the same aristotype cubic structure as conventional inorganic perovskites but incorporate at the A-site an organic molecule, possibly polar and which can potentially rotate. Due to their appropriate bandgap \cite{Quarti2016,Ziffer}, high absorption coefficients \cite{Ball} and exceptionally long carrier lifetimes and diffusion lengths \cite{Stranks,Wehrenfennig,Xing344}, OMHPs, and in particular lead-based compound like methylammonium lead iodide (CH$_{3}$NH$_{3}$PbI$_{3}$, or MAPbI$_{3}$), produced recently a revolution in the field of photovoltaics \cite{mcgehee2014,yang2017,zheng2020}. Despite the tremendous progress regarding device efficiency, the physical origin of the extraordinary performance of MAPbI$_{3}$ solar cells is still poorly understood. Ferroelectric (FE) domains, acting as internal p-n junctions that efficiently separate and transport the photoexcited electron-hole pairs \cite{grinberg,frost2014nano,liu2015}, are believed to be an important factor explaining their efficiency.

Definitive consensus and explanation regarding the FE character of MAPbI$_{3}$ and related perovskites are however still missing \cite{shahrokhi2020}. On the one hand, experimental fingerprints of ferroelectricity were reported in single crystals, relying on quasielastic neutron scattering \cite{leguy2015} or dielectric measurements \cite{juarez2014,anusca2017}. Some FE hysteresis loops were also measured at low temperature (200K) \cite{dong2016,rakita2017}. Nevertheless, no clean spontaneous polarization measurement has been reported so far for MAPbI$_{3}$ at room temperature, due to the relatively high electrical conductance and low stability under high applied voltage bias \cite{zhang2018}. FE/ferroelastic domains were observed in the tetragonal phase of MAPbI$_{3}$ polycrystalline films \cite{rohm2017,kutes2014,liu2018} as well. On the other hand, some independent reports  \cite{gomez2019,frohna2018,weller2015,kawamura2002} questioned or even excluded the possibility of a FE character which remains therefore under hot debate. 

Most popular ferroelectrics, like BaTiO$_3$ or PbTiO$_3$, belong to the family of inorganic perovskites, suggesting that ferroelectricity might be inherent to the perovskite structure. However, it is worth noticing that most ABO$_3$ perovskites with a small Goldschmidt tolerance factor $t$  \cite{goldschmidt} are not FE due to a typical competition between polar lattice distortions and antiferrodistortive (AFD) rotation motions of the BO$_6$ octahedra \cite{benedek2013,amisi2012}. The latter motions usually dominate for $t< 1$ in ABO$_3$ compounds and suppress the FE instability,  giving rise to a common non-polar $Pnma$ ground state that arises from the joint condensation of in-phase and out-of-phase rotation motions, according to an $a^-b^+a^-$ rotation pattern in Glazer's notations \cite{glazer1972}.

With its small tolerance factor ($t= 0.912$~\cite{baikie}), MAPbI$_{3}$  seems perfectly in line with this generic behavior. Stable at high-temperature in a cubic $Pm\bar{3}m$ phase, it exhibits on cooling two consecutive phase transitions arising from the condensation of AFD distortions: a first transition at 327 K to a tetragonal phase {\it a priori} of $I4/mcm$ symmetry showing an AFD distortion pattern $a^0a^0c^-$ and a second transition at 162K to an orthorhombic ground state of $Pnma$ symmetry with the typical distortion pattern $a^-b^+a^-$. As such, MAPbI$_{3}$ seems to behave like a conventional non-FE perovskite. However, while there is a broad consensus on the non-polar $Pnma$ character of the ground state \cite{weller2015,lee2015cc,lee2016cm}, the exact assignment of the tetragonal phase to either the non-polar $I4/mcm$ or instead a polar $I4cm$ space group with potential FE character is still unclear.

\begin{figure*}
\includegraphics[width=\linewidth]{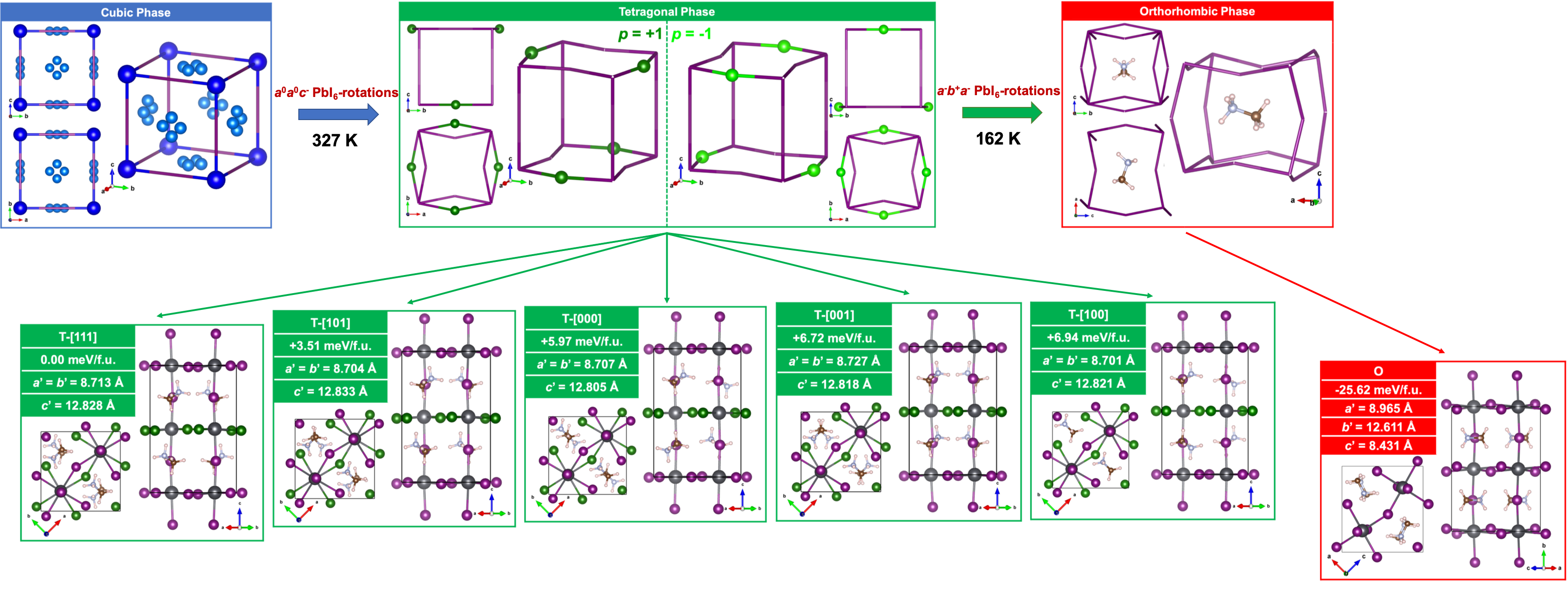}
 \caption{\textbf{Preferred orientations of MA$^+$ molecules under different phases.} Herein, balls represent the projection of the C-N bond axis from C to N atom. In the cubic phase, 8 blue balls and 24 light-blue balls correspond to $<$111$>$- and $<$100$>$-oriented cations, respectively. For the tetragonal phase, preferred orientations of the organic ions reduce to four (green balls), lying in the $<$101$>$ directions. When it goes to the orthorhombic phase, the orientation of MA$^+$ molecules is fixed. Tetragonal phases in the same energy region within various polarization states, as well as the antiferroelectric orthorhombic phase are displayed in the bottom part. Their lattice constants are given in the basis of corresponding superlattices. Note that different tetragonal phases are entitled according to their net polarization directions. Molecules of all of them lie in $<$101$>$.}
 \label{fig1}
\end{figure*}

A key distinction between inorganic ABO$_3$ perovskites and MAPbI$_{3}$ is the presence of polar molecules at the A site, which introduces the possibility of an order-disorder FE transition like in PVDF \cite{PVDF1983}. As anticipated in previous studies~\cite{stroppa2015}, not only the displacive mechanism of inorganic perovskites (polar soft mode) but also the order-disorder arrangement of the polar MA$^{+}$ molecule can contribute and collaborate to make MAPbI$_3$ FE. Several studies have highlighted the complexity of MA$^{+}$ arrangement \cite{mosconi2014,quarti2014,leguy2015,kulbak2015,lee2016sr,bechtel2016,egger2016,brenner2016,frost2016,zhu2017} and its important role on controlling the electronic/optical properties of MAPbI$_{3}$ \cite{mosconi2013,amat2014,quarti2014,frost2014nano,zheng2015,fan2015,xiao2015,motta2015}. Moreover, MA$^{+}$ dipole direction is believed to closely correlate with AFD rotations and tilts of PbI$_{6}$-octahedra through hydrogen bonding  \cite{quarti2015,lee2015cc,li2016,lee2016cm,lee2016sr}. Exploring how the complicated interplay between MA$^{+}$ molecules and the PbI$_{6}$-octahedra network might impact FE properties is thus essential. However, no global picture describing the evolution of molecule orientations, polar distortion and AFD motions in all three phases has been reported yet.

Here, using first-principles density functional theory (DFT) calculations, we systematically investigate the FE properties in MAPbI$_{3}$ going step by step from the cubic to the tetragonal and then orthorhombic phases. We rationalize the impact of AFD distortions on the preferential orientations of the MA$^{+}$ molecules. From this, we explain the non-FE character of the orthorhombic phase and to which extent the tetragonal phase can be considered as FE. Our work intents to reconcile previous observations, providing a simple unified and comprehensive picture regarding the influence of AFD motions on the potential FE character of MAPbI$_{3}$.

\section*{RESULTS}

\noindent
\textbf{Cubic Phase.} MAPbI$_3$ is a typical ABX$_3$ perovskite but that includes at the A site an ionized molecule MA$^+$ which, according to the difference of electronegativity of C and N atoms, is moreover polar (with its dipole moment pointing from C to N). This polar molecule breaks by itself the symmetry of the aristotype cubic perovskite structure. At high temperature, the MA molecule can however rotate relatively freely and MAPbI$_3$ is reported in a cubic $Pm \bar{3}m$ symmetry \cite{onoda1990,weller2015}. This is nevertheless an average symmetry resulting from the random orientation of the MA molecules. The disordered and dynamical character of the $Pm \bar{3}m$ phase makes it more hardly accessible from static DFT calculations with periodic boundary conditions than other static cases.

\begin{figure}
\centering
 \includegraphics[width=\linewidth]{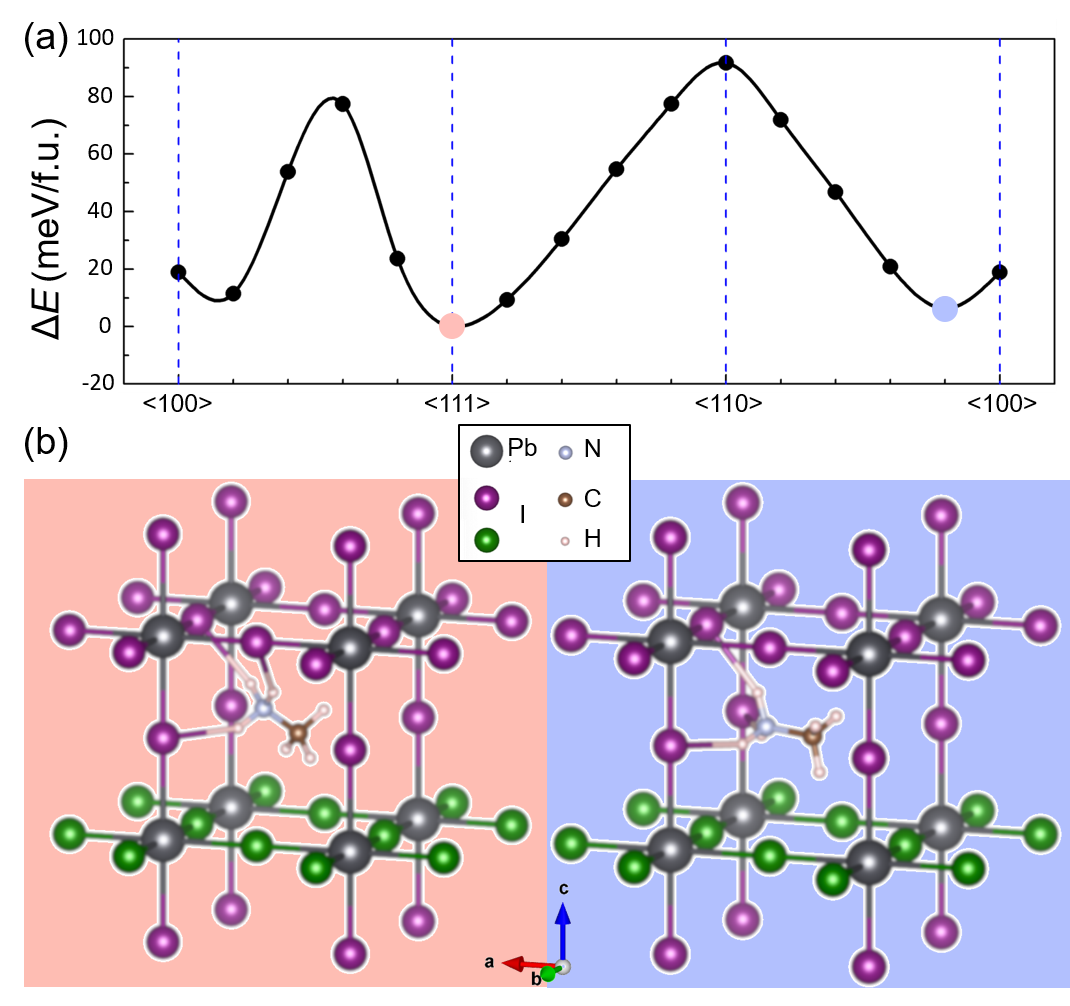}
 \caption{\textbf{MA$^+$ molecule's orientation in cubic phase.} (a) Evolution of energy when the axis of MA$^+$ molecules is oriented along different principal directions ($1 \times 1 \times 1$ cubic cell). The lowest-energy configuration with $<$111$>$-oriented molecules is taken as energy reference. (b) Sketch of the two cubic states of lowest energies. Molecules in the left and right configurations are pointing toward $[111]$ and $[100]$ directions respectively. The three bonds formed between nitrogen-end-type hydrogen atoms and iodine atoms are displayed.}
 \label{fig2}
\end{figure}

In order to investigate the energy landscape within which the MA molecule is rotating and its preferential orientations, we consider first a $1 \times 1\times 1$ cubic cell with Pb and I atoms fixed at high-symmetry positions, and relaxed the molecule position for various orientations.  The results are illustrated in Fig. \ref{fig2}(a), summarizing the evolution of the energy for the C-N axis of the molecule rotating from the $<$100$>$ to the $<$110$>$ and then the $<$111$>$ directions of the cubic unit cell. Consistently with previous reports \cite{motta2015,li2016}, we observe two close local energy minima $E_{<100>}$ and $E_{<111>}$ associated with preferential orientations of the MA$^+$ molecule along the $<$100$>$ and $<$111$>$ directions, respectively. The optimized lattice parameters are 6.317 \AA\ (6.305 \AA) for the $<$111$>$ ($<$100$>$) state, in good agreement with the experimental value of 6.31 \AA\ \cite{stoumpos2013}. Similarly to Yin's calculations \cite{yin2015}, $E_{<111>}$ is $\sim$ 6 meV per formula unit (f.u.) lower than $E_{<100>}$. However, the relative stability of both states can be reversed when properly relaxing the octahedra cage ($E_{<111>}-E_{<100>}= 12$ meV/f.u.), in line with what was observed in some alternative studies \cite{brivio2013,frost2014apl,egger2014}. In practice, these states can be considered as nearly energetically equivalent.

\begin{figure}
\includegraphics[width=0.5\linewidth]{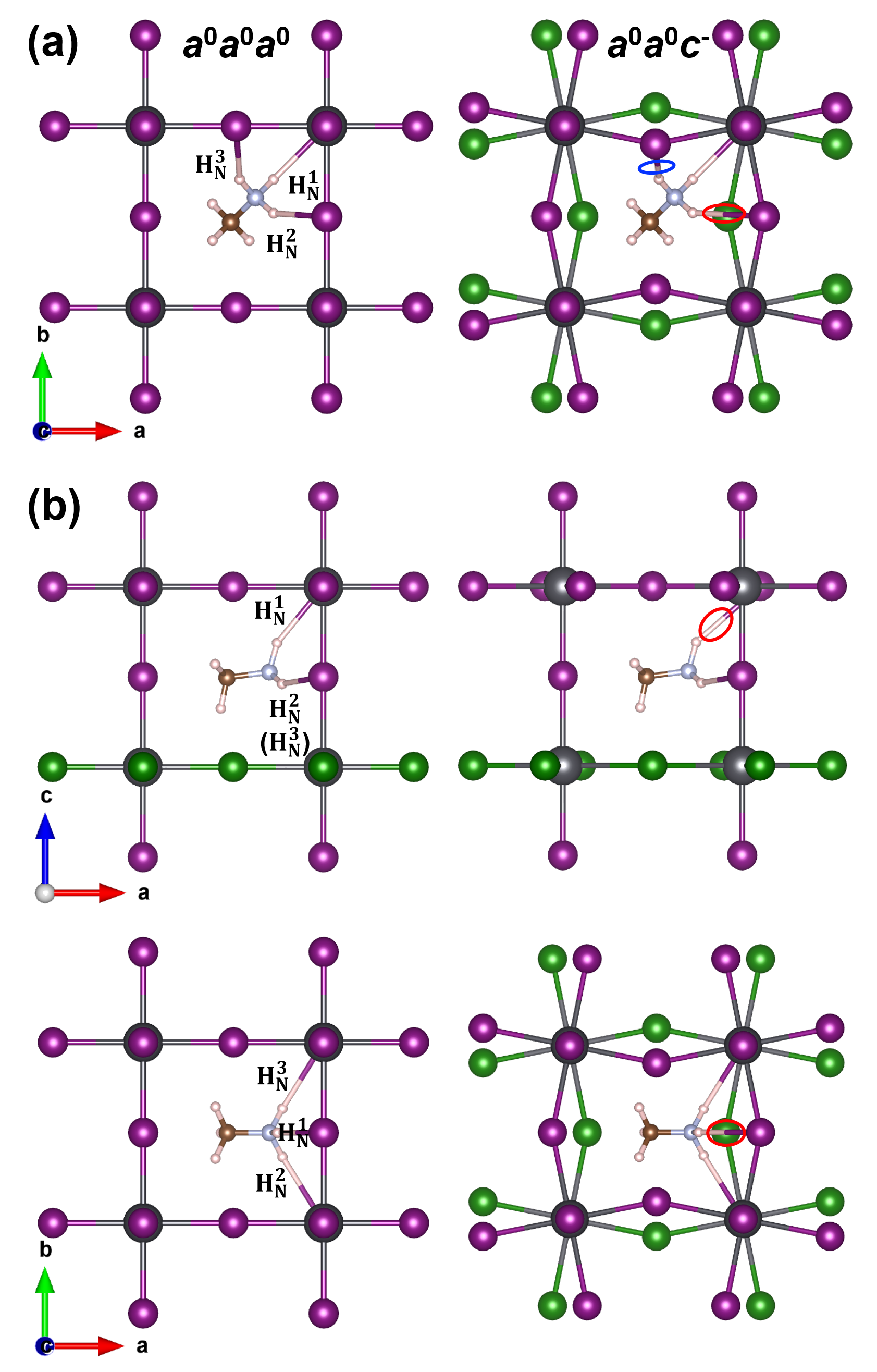}
 \caption{\textbf{The role of PbI$_{6}$-octahedra rotations on reorientation of MA$^+$ molecules from cubic to tetragonal phase.} For the cases with (a) $[111]$- and (b) $[100]$-oriented molecule, stretched (red) and compressed (blue) H$_\mathrm{N}$-I bonds due to $a^0a^0c^-$ PbI$_{6}$-octahedra rotations are circled. The octahedron rotation factor $p = -1$ here.}
 \label{fig3}
\end{figure}

The atomic configurations associated with the $E_{<111>}$ and $E_{<100>}$ minima are represented in Fig. \ref{fig2}(b). Both the off-centering of the molecule and its slight angular deviation from the ideal $[100]$ direction in the $<$100$>$ state  (see right panel of Fig. \ref{fig2}(b) and up panels of Fig. \ref{fig3}(b)) highlight the formation of hydrogen bonds between the molecule and the inorganic backbone \cite{lee2015cc,lee2016cm,lee2016sr}, i.e. bonds between the three nitrogen-end-type hydrogen atoms (H$_\mathrm{N}$) and iodine atoms, as well as the trend to make these three H$_\mathrm{N}$-I bonds equivalent. We notice that properly reproducing this in the calculation is not necessarily straightforward and requires some special care during the atomic relaxation (i.e. properly rotating the molecule about the C-N axis). As summarized in Fig. \ref{fig1}, the molecule will therefore preferably align along one the of eight $<$111$>$ directions or along one of the six $<$100$>$ directions with, in that case, four possible distinct configurations, associated with small canting driven by H$_\mathrm{N}$ atoms bonding with different I.

The previous results are obtained using a $1 \times 1 \times 1$ periodic cell so that all periodically-repeated molecules are artificially aligned. In such a case, the structure is therefore naturally polar and shows very similar spontaneous polarizations of 9.79 and 9.91 $\mu$C/cm$^2$ in the $<$111$>$ and $<$100$>$ states respectively. This spontaneous polarization can be seen as the sum of two contributions: the first one, labelled ``order-disorder'' ($P_\mathrm{o-d}$), arises from the alignment of the intrinsic dipole moment of the MA$^+$ cations, while the second one,  labelled ``displacive'' ($P_\mathrm{dis}$), comes from the relative polar motion of cations and anions respect to centrosymmetric positions. $P_\mathrm{o-d}$ is estimated to 5.66 (5.37) $\mu$C/cm$^2$ for the molecule along $<$111$>$ ($<$100$>$) direction, a value comparable to that of 5.56 $\mu$C/cm$^2$ reported by Stroppa \textit{et al.}\cite{stroppa2015} and roughly 50\% of the total polarization ($P_\mathrm{tot}$). This highlights that both order-disorder (i.e. alignment of MA molecule) and displacive (i.e. polar soft mode) mechanisms equally contribute to produce a polarization when enforcing a polar phase in cubic MAPbI$_3$ but this does not guarantee that such a state will naturally emerge.

Considering larger supercells and orienting differently the molecules from site to site either along the $<$111$>$ or $<$100$>$ directions, we can generate various non-polar configurations that appear at energies similar to the polar one, as long as neighboring molecules don't try to bond with the same iodine atoms. This means that, except through their interactions with I atoms, the molecules behave rather independently and there is no clear tendency for them to align from site to site. Looking more globally at the energy landscape in Fig.  \ref{fig2}(a), it appears relatively flat, in agreement with previous calculations \cite{brivio2013}. The energy barriers between distinct minima are small ensuring that the molecule can freely rotate at sufficiently high temperature \cite{wasylishen1985}. All this supports the view of a disordered cubic phase in which molecules rotate rather independently and without strong tendency to align and form a polar state.

\

\noindent
\textbf{Tetragonal Phase.} Consistently with its small tolerance factor, MAPbI$_{3}$  undergoes on cooling slightly above room temperature ($T = 327$ K) a first structural phase transition arising from the condensation of AFD rotations of PbI$_{6}$-octahedra according to the $a^0a^0c^-$ pattern. The emergence of these rotations lowers the symmetry of the system from $Pm\bar{3}m$ to $I4/mcm$. This distorts the originally cubic environment at the A site and is therefore expected to affect the behavior of the MA$^+$ molecules \cite{li2016,lahnsteiner2016}. In inorganic perovskites, octahedra rotations typically reduce the natural tendency to polar distortion and emergence of displacive ferroelectricity. It is however questionable if it could here force the polar molecules to align and favor instead order-disorder ferroelectricity.

As highlighted in the previous Section, the preferential orientations of the MA$^+$ molecules are mainly driven by the formation of three equivalent hydrogen bonds. In Fig. \ref{fig3}, we illustrate how the bonds are affected by the octahedra rotations. When the molecule is aligned along the $[111]$ direction, as sketched in Fig. \ref{fig3}(a), anticlockwise AFD motions typically induce a compression of one H$_\mathrm{N}$-I bond and the stretching of another one, forcing the molecule to rotate from the $[111]$ to the $[101]$ direction. This reorientation is generic to any $<$111$>$ directions although the specific final orientation of the molecule [$a', b', c'$] depends on both its initial direction [$a, b, c$] ($a,b,c = \pm 1$) and the clockwise or anticlockwise character $p$ of the octahedron rotation ($p = \pm 1$) according to the following relationship:
\begin{equation}
\begin{pmatrix}
a'\\[1.2ex]
b'\\[1.2ex]
c' 
\end{pmatrix}  
= 
\begin{pmatrix}
\frac{1-cp}{2} & 0 & 0\\[1.2ex]
0 & \frac{1+cp}{2} & 0\\[1.2ex]
0 & 0 & 1 
\end{pmatrix}   
\begin{pmatrix}
a\\[1.2ex]
b\\[1.2ex]
c 
\end{pmatrix}
\end{equation}

When the molecule is aligned instead in the $<$100$>$ directions, it prefers to stay perpendicular to the axis of PbI$_{6}$-rotations. Then, as shown in Fig. \ref{fig3}(b), appearance of PbI$_{6}$-rotations stretch one H$_\mathrm{N}$-I bond forcing the MA$^{+}$ molecule to rotate from the $[100]$ direction to the $[101]$ direction again. This time, the final direction of the molecule can be obtained as $[ \pm \frac{k-1}{2}, \pm \frac{k+1}{2}, kp ]$ and depends of its original direction $[ \pm \frac{k-1}{2}, \pm \frac{k+1}{2}, 0 ]$ ($k = \pm 1$) and the character of the octahedron rotation $(p = \pm 1$).

This justifies that, related to hydrogen bonding effects, the $a^0a^0c^-$ PbI$_{6}$-rotations in the tetragonal phase produce the reorientation of MA$^+$ cations along $<$101$>$ directions, in agreement with experimental X-ray diffraction measurements on single crystals \cite{kawamura2002}. However, the previous discussion points out that all $<$101$>$ orientations are not equivalent. As illustrated in Fig. \ref{fig1}, a distinct set of four orientations will be favored according to the distortion produced by the local clockwise ($p=1$; dark green) or anticlockwise ($p=-1$; light green) character of the PbI$_{6}$-rotations. This means that the appearance of the $a^0a^0c^-$ AFD rotation pattern, which alternates clockwise and anticlockwise rotations from site to site along the three directions, will prevent the development of a polar state with all molecules aligned along the same direction and will favor instead configurations in which molecules at neighboring sites are either perpendicularly oriented or anti-aligned (populating alternatively light and dark green directions in Fig. \ref{fig1}).

In order to further assess the validity of this, we consider $\sqrt{2}\times\sqrt{2}\times2$ supercells with various possible orientations of the four independent molecules. All identified low-energy structures correspond to configurations with molecules at neighboring sites either perpendicular or anti-aligned as illustrated in Fig. \ref{fig1}. According to the relative orientations of the molecules, some of these structures are polar and can develop a polarization along distinct directions. Due to partial compensation of dipole moments from cell to cell, the polarization is reduced compared to the case with all molecules aligned but it includes again almost equal $P_\mathrm{dis}$ and $P_\mathrm{o-d}$ contributions and the latter can be estimated from the orientation of a rigid dipole moment associated with each molecule (see Supplementary Material Table. SI for the polarization details of tetragonal phases). 

The lowest energy configuration that we found corresponds to a polar state T-[111] with all molecules sharing the component along the same direction of $c$-axis and giving rise to a net polarization of 7.74 $\mu$C/cm$^2$ oriented nearly along the [111] direction. But various other polar states -- some of them previously discussed in the literature -- with their net polarization along either the [101] direction \cite{frost2014nano,brivio2015} (+3.51 meV/f.u., T-[101]) or the [001] direction \cite{quarti2014,stroppa2015} (+6.72 meV/f.u., T-[001) or the [100] direction (+6.94 meV/f.u., T-[100]) and even an antipolar state (+5.97 meV/f.u., T-[000]) appear very close in energy. So, although the lowest energy phase is polar, this highlights a energy landscape with many local minima almost at the same energy, compatible with the view of a partly dynamical tertagonal phase at room temperature \cite{chen2015,leguy2015}, with molecules hopping from one direction to another and no very strong tendency to ferroelectricity again.

\begin{figure}
 \includegraphics[width=\linewidth]{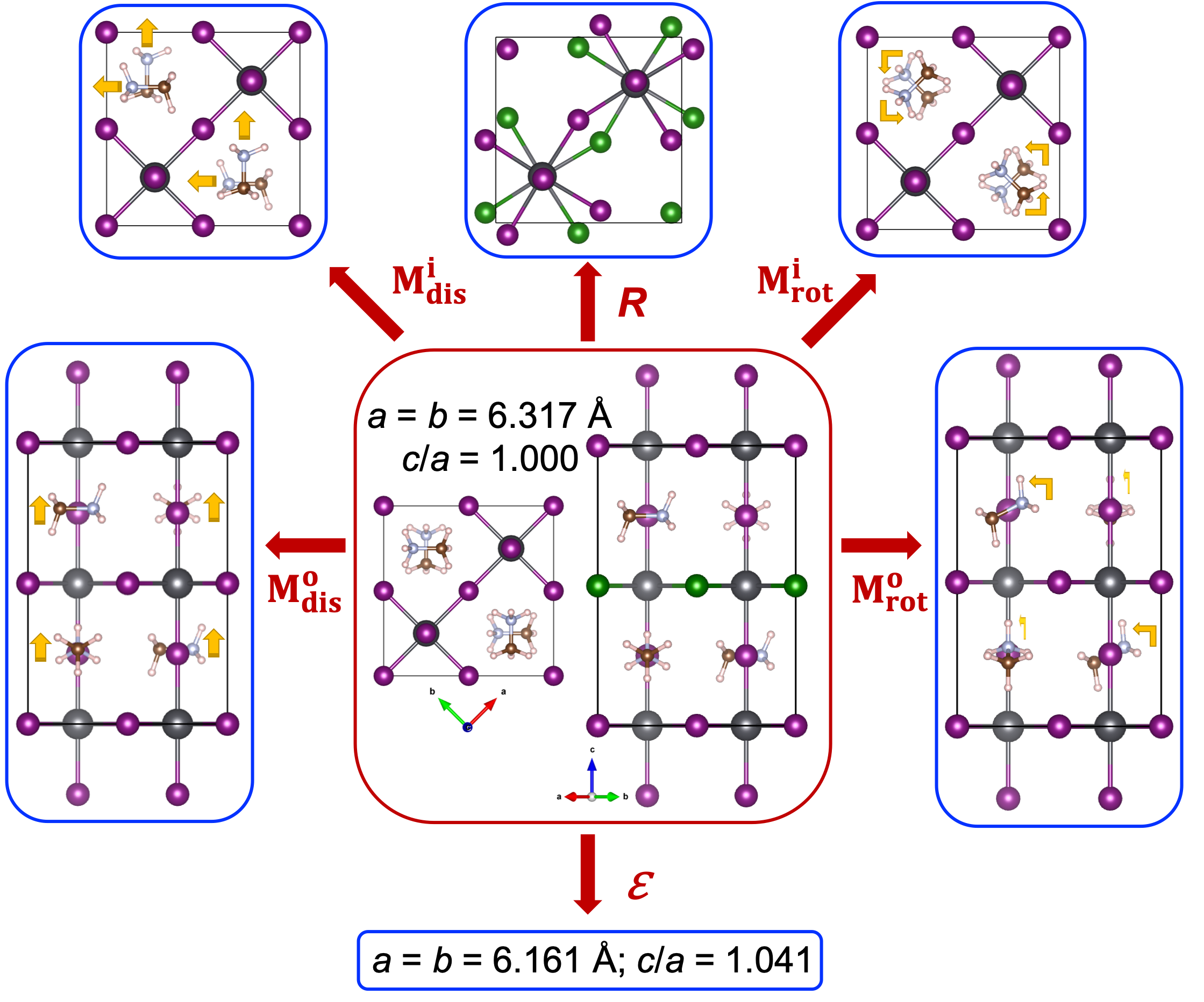}
 \caption{\textbf{Sketch of six structural degrees of freedom considered in our energy analysis.}}
 \label{fig4}
\end{figure}

To better clarify the interplay between distinct structural distortions, we investigate how the energy is evolving from an artificial ``cubic'' phase (C) taken as reference to the lowest-energy polar tetragonal T-[111] phase. For the reference phase, the unit cell is fixed cubic, the Pb and I atoms as well as the center of mass of the MA$^+$ molecules are located at high-symmetry positions and the MA$^+$ molecules oriented each along the [110]-type direction that is the closest to the corresponding direction in the T-[111] phase. Strictly speaking, symmetry of this structure is not cubic. It even shows (see the center of Fig. \ref{fig4}) a net polarization along the [010] direction but constitutes a suitable reference for our purpose. The total distortion from this reference to the T-[111] phase is then spread into six independent contributions illustrated in Fig. \ref{fig4}: the tetragonal macroscopic strain relaxation of the unit cell ($\varepsilon$), the a$^0$a$^0$c$^-$ rotation of the PbI$_{6}$ octahedra ($R$), rotations of MA$^+$ molecules about either in (M$^\mathrm{i}_\mathrm{rot}$) or out-of (M$^\mathrm{o}_\mathrm{rot}$) $a$-$b$ plane and displacive polar motions in (M$^\mathrm{i}_\mathrm{dis}$) or out-of (M$^\mathrm{o}_\mathrm{dis}$) $a$-$b$ plane. Then a simple model energy is written as:
\begin{equation}
E(\mathrm{T}) = E(\mathrm{C}) + \sum_{i=1}^{6} F_i+\sum_{i \neq j} F_{ij} + \sum_{i \neq j \neq k} F_{ijk}
\end{equation}
with $i, j, k$ running over the six distortions and $F$ energy contributions fitted separately on first-principles calculations including related individual and combined distortions. The deviation of energy between first principles and the calculated $E$(T-[111]) from Eq. (2) is less than 0.03\%, proving its reliability.
%reproduce $E(T_{[111]})$, not included in the fit process, with an accuracy of 0.03\%.

\begin{table}
    \centering
\begin{center}
    \begin{tabular}{c|cccccc}
    \multicolumn{7}{c}{$F_i$ and $F_{ij}$} \\
    \multicolumn{7}{c}{ }\\
 \hline
    \hline
        \diagbox{\it{j}}{\it{i}} & $\varepsilon$ & $R$ & M$^\mathrm{o}_\mathrm{dis}$ & M$^\mathrm{o}_\mathrm{rot}$ & M$^\mathrm{i}_\mathrm{dis}$ & M$^\mathrm{i}_\mathrm{rot}$ \\
    \hline
       $\varepsilon$ & 46.12 & {\color{red}{\textbf{-79.17}}} & 2.88 & 1.69 & 27.13 & -3.44\\
       $R$ & & -14.79 & 53.42 & 35.95 & 63.40 & {\color{red}{\textbf{-92.20}}} \\ 
       M$^\mathrm{o}_\mathrm{dis}$ & & & 
       \cellcolor{lightgray} 
       -8.49 & \cellcolor{lightgray} {\color{red}{\textbf{-39.43}}} & \cellcolor{lightgray} 13.75 & \cellcolor{lightgray} 11.21 \\ 
       M$^\mathrm{o}_\mathrm{rot}$ & & & & 
       \cellcolor{lightgray} -5.23 & \cellcolor{lightgray} 18.19 & \cellcolor{lightgray} -2.72 \\ 
       M$^\mathrm{i}_\mathrm{dis}$ &&&&& 
       \cellcolor{lightgray} {\color{red}{\textbf{-79.23}}} & \cellcolor{lightgray} {\color{red}{\textbf{-46.59}}}\\
       M$^\mathrm{i}_\mathrm{rot}$ &&&&&& 
       \cellcolor{lightgray} -5.11\\ 
     \hline
     \hline
\end{tabular}
\end{center}

    \caption{\textbf{Values (meV/f.u.) of the fitted $F_i$ and the $F_{ij}$ terms.} Most significant negative terms are highlighted in red. Light-gray-highlighted data are in the cubic regime without rotations.}
    \label{table1}
\end{table}

The first- and second-order $F$ values are reported in TABLE \ref{table1}. Starting from the C phase and looking at $F_i$ contributions, the dominant term to decrease the energy is $F$[M$^\mathrm{i}_\mathrm{dis}$] (-79.23 meV/f.u.) that corresponds to the off-centering of the MA$^+$ molecules  driven by the formation of H$_\mathrm{N}$-I bonds as previously discussed and really existing in the $<$110$>$ case in Fig. \ref{fig2}(a). It appears here as a polar motion because the molecule arrangement in the C phase is already polar but it would take the form of an antipolar distortion in a C phase with antipolar alignment of the molecules.

\begin{figure}
 \includegraphics[width=\linewidth]{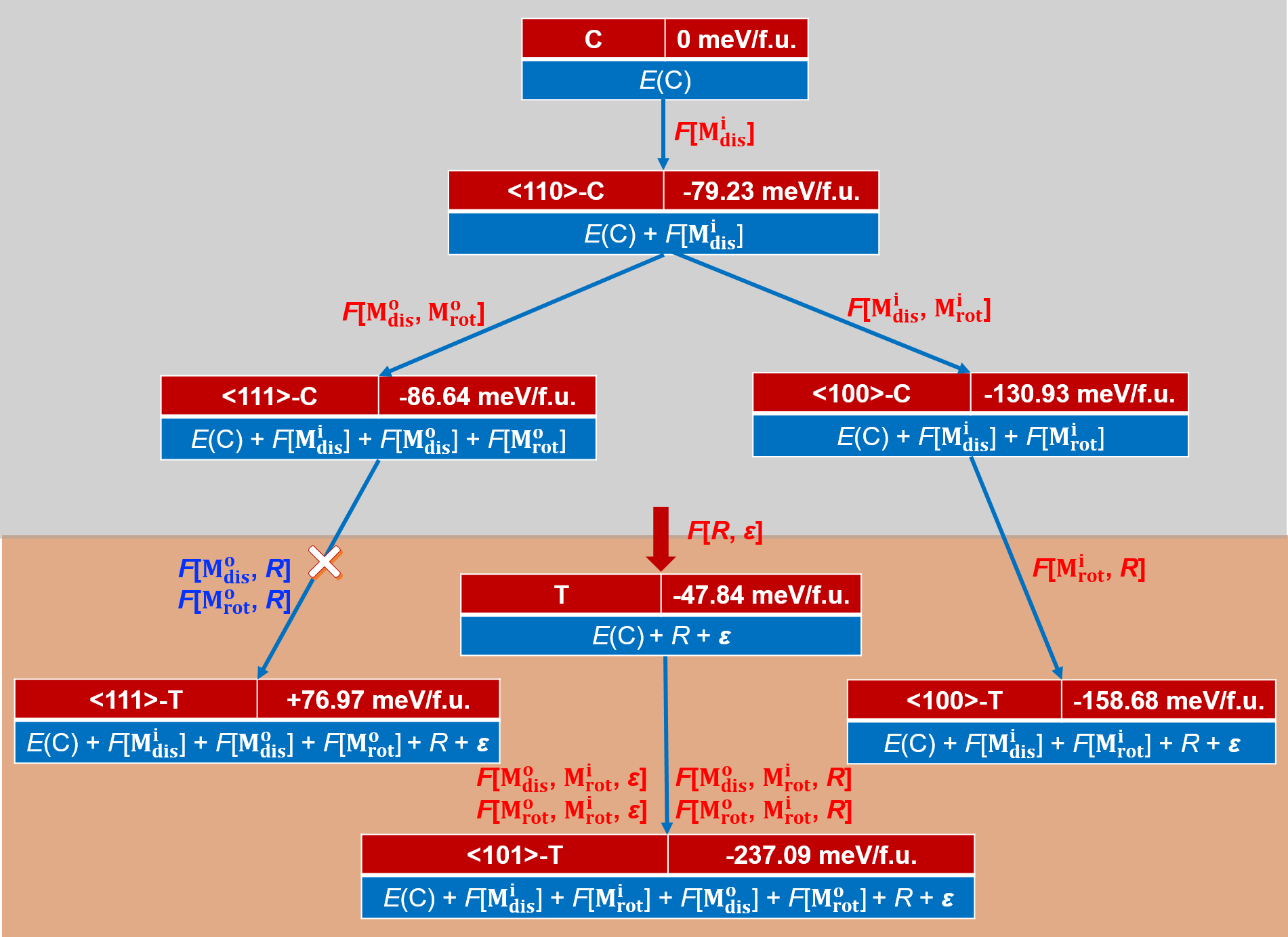}
 \caption{\textbf{Connection between the reference C phase and the T-[111] phase.} When including successively the distinct degrees of freedom listed in Fig. \ref{fig4}, the reference C phase with $<$110$>$-oriented molecules gradually transforms to the T-[111] lowest-energy phase with $<$101$>$-oriented molecules. Light-gray and light-orange backgrounds refer to phases without and with $R$ distortion. Dominant negative and positive energy contributions are highlighted respectively in red and blue.}
 \label{fig5}
\end{figure}

Ignoring at first octahedra rotations ($R$), we notice the strong cooperative coupling of M$^\mathrm{i}_\mathrm{dis}$ with M$^\mathrm{i}_\mathrm{rot}$ that will produce a reorientation of the molecules along the [100]-type directions ($<$100$>$-C phase in Fig. \ref{fig5}) in agreement with the related minimum in Fig. \ref{fig2}(a). We can obtain another low-energy configuration for the molecules along the [111]-type directions ($<$111$>$-C phase in Fig. \ref{fig5}),  but our model with few rigid degrees of freedom connected to the ground state is not appropriate for full relaxations and this requires considering only the part of M$^\mathrm{i}_\mathrm{dis}$ related to the translation along the molecule. All these can be interpreted as molecular motions driven by H-bonding formation.

Then, $F[R]$ is negative in Table \ref{table1} (-14.79 meV/f.u.) highlighting that AFD rotations of PbI$_6$ octahedra are spontaneously unstable, in line with the small tolerance factor of MAPbI$_3$. We first observe a strong coupling of these rotations with the macroscopic tetragonal strain that significantly helps stabilizing an AFD phase ($F[R, \varepsilon]$= -79.17 meV/f.u.). Further the significant coupling of $R$ and M$^\mathrm{i}_\mathrm{rot}$ strongly favors orientation of the molecules along the [100]-type rather than the [111]-type directions. Interestingly, while $F$[M$^\mathrm{i}_\mathrm{dis}$, M$^\mathrm{i}_\mathrm{rot}$] and $F$[M$^\mathrm{o}_\mathrm{dis}$, M$^\mathrm{o}_\mathrm{rot}$] appear globally competitive in absence of octahedra rotations (favoring molecule orientations along either [100]- or [111]-type directions), thanks to different third-order terms listed in Supplementary Material Table SII(a), they become cooperative in presence of $R$ and $\varepsilon$ and molecules thus favor [101]-type orientations ($<$101$>$-T phase in Fig. \ref{fig5}). To explore the driving force for the cooperation, we fit some related fourth-order terms that are not taken into account in Eq. (2). From Supplementary Material Table SII(b), the terms within $\varepsilon$ trend to increase the total energy, which nicely confirms that the reorientation of the molecules along the $<$101$>$ directions in the tetragonal phase results from the AFD motions. Note that, this appears mainly driven by local steric and geometric effects compatible with a partial disorder from site to site. The combination of distortions gives rise here to the T-[111] lowest energy phase but a very similar decomposition can be obtained for the other T phases in Fig. \ref{fig1} highlighting no strong tendency to favor a particular polar alignment.

\

\noindent
\textbf{Orthorhombic Phase.} Below 162 K, MAPbI$_{3}$ adopts an orthorhombic $Pnma$ structure with a PbI$_6$ rotation pattern $a^-b^+a^-$. Comparing to the tetragonal phase including a single out-of-phase rotation about $c$-axis, this arises from the appearance of two more AFD distortions: out-of-phase rotations about $a$-axis and in-phase rotations about $b$-axis. Our calculations confirm that this phase is significantly lower in energy than the tetragonal phases and the proper ground state of MAPbI$_{3}$. Similar to what happens in the tetragonal phase, the two new AFD distortions will further constraint the direction of the MA$^+$ molecules.

\begin{figure}
\includegraphics[width=\linewidth]{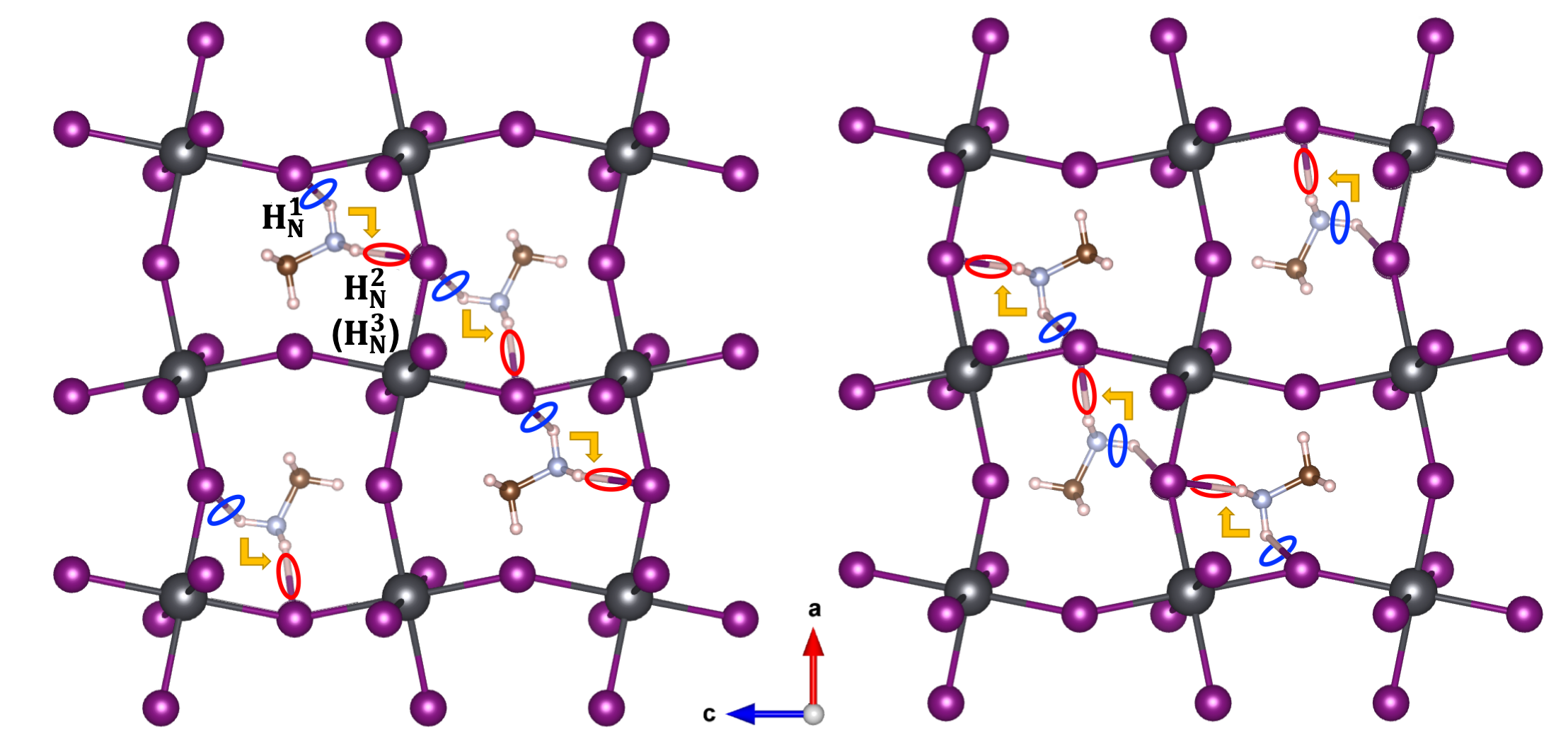}
 \caption{\textbf{Atomic configuration in the $a$-$c$ plane for two consecutive layers along $b$-axis in the $a^-b^+a^-$ orthorhombic phase.}  MA$^+$ molecules are displayed according to their expected original positions (along $[10\overline{1}]$ or $[\overline{1}01]$ directions) in the phase with $a^-b^0a^-$ rotation pattern. Presence of the additional $a^0b^+a^0$ distortion then produces the compression (blue circles) and stretching (red circles) of some H$_\mathrm{N}$-I bonds and forces molecules‘ reorientation according to the yellow arrows. The final structure is necessarily antipolar.}
 \label{fig6}
\end{figure}

In the tetragonal phase, we have seen that the out-of phase rotations along $c$-axis impose the MA$^+$ molecule axis to be aligned along generic directions defined as $[ \pm \frac{k-1}{2}, \pm \frac{k+1}{2}, kp ]$. Similarly, the appearance of a new out-of-phase rotation along $a$-axis will force alignment along $[k'p, \pm \frac{k'-1}{2}, \pm \frac{k'+1}{2}]$ directions (taking, without loss of generality, the convention that $p$ is the same for both rotations in a given cell). Combining these two constraints, the only possibility is $k=-1$ and $k'=1$ so that the molecules can only be aligned along generic directions $[p, 0, -p]$, with $p$ switching from $+1$ to $-1$ when moving from one unit cell to the next one along $a$, $b$ or $c$ direction. This gives rise to an antipolar configuration in which molecules at neighboring sites are oppositely aligned (see Fig. \ref{fig6}). 

The appearance of the additional in-phase rotations along $b$-axis will further distort the structure, contracting the H$^1_{\rm N}$-I bond (blue-circled in Fig. \ref{fig6}) and expanding the H$^2_{\rm N}$-I and H$^3_{\rm N}$-I bonds (red-circled in Fig. \ref{fig6}). This will force the molecules to rotate nearly 45$^{\circ}$ either clockwise or anti-clockwise and align their axis closely along $<$100$>$ directions. Due to their reversed rotations, molecules at neighboring sites in the same $a$-$c$ plane are perpendicularly oriented. Taking into account the in-phase character of the rotations along $b$-axis, the molecules will adopt in consecutive $a$-$c$ planes generic directions corresponding respectively to $[\frac{p-1}{2}, 0,  -\frac{p+1}{2}]$ and $[\frac{p+1}{2}, 0, -\frac{p-1}{2}]$. This is associated with a pattern in which first-neighbor molecules along $a$ and $c$ are perpendicularly aligned (second neighbors being aligned) while first-neighbor molecules along $b$-axis are anti-aligned.  

In the $Pbnm$ phase, due to the combination of the AFD motions, the MA$^+$ molecules are therefore forced to take an antipolar arrangement in which planes with opposite net polarizations in $[101]$ and $[\overline{1}0\overline{1}]$ directions alternating along $b$-axis. This demonstrates unambiguously that, in line with the literature \cite{weller2015,lee2015cc,lee2016cm}, this phase is antipolar and cannot be FE.

\section*{DISCUSSION}
\noindent
In contrast to inorganic ABO$_3$ perovskites, MAPbI$_3$ possesses MA$^+$ polar ions at the A site. As such, its FE character, rather than displacive, is closely tied to the  arrangement of these permanent dipoles, which is itself determined by the interplay of two main phenomena. On the one hand, the local positions and orientations of MA$^+$ dipoles are guided by the natural tendency of nitrogen-end-type hydrogen atoms to form three equivalent H$_\mathrm{N}$-I bonds with the surrounding inorganic lattice. On the other hand, like traditional inorganic ABO$_3$ perovskites with small tolerance factors, MaPbI$_3$ shows AFD octahedra-rotation instabilities that give rise to consecutive phase transitions from cubic to $a^0a^0c^-$ tetragonal and then $a^-b^+a^-$ orthorhombic distorted phases. These cooperative AFD distortions modulate the shape of the A-site cavities forcing then some specific orientations of the molecules at each atomic site.  

At high temperature, in absence of AFD distortions the A-site cavities keep a cubic symmetry. Due to their tendency to H-bonding, the molecules show preferential orientations along either $<$100$>$ or $<$111$>$ directions but the energy landscape is sufficiently flat for the molecules to rotate relatively freely and independently as typically observed experimentally \cite{wasylishen1985} and also confirmed from molecular dynamics simulations \cite{mosconi2014,frost2014apl}. This gives rise to a dynamically-disordered non-polar average structure which can be seen as a disordered paraelectric state.

At $T = 327$ K, MaPbI$_3$ then evolves to a tetragonal $I4/mcm$ phase arising from the condensation of a $a^0a^0c^-$ pattern of octahedra rotations. These AFD motions distort the A-site cavities forcing a partial reorientation of the MA$^+$ molecules along the $<$101$>$ directions, although leaving them some degrees of freedom. At a given site, four preferential orientations perpendicular to each others are allowed while at neighboring sites they are either perpendicular or anti-aligned. In this tetragonal phase, the T-[111] structure owing the lowest energy is polar, with a spontaneous polarization arising evenly from the partial alignment of the MA$^+$ dipoles and from the further polarizability of the lattice. This latter effect can however be seen as a pseudo-proper effect rather than a driving force for the appearance of the polarization since it would translate into an antipolar contribution for antipolar molecular arrangements. Consequently to the presence of PbI$_6$ rotations, this polar state does not correspond to a na\"{i}ve configuration of perfectly aligned dipoles but is one in which MA$^+$ molecules at neighboring sites are perpendicularly oriented giving rise to what we could call a weakly polar configuration. According to this, the polarization reversal is possible so that the T-[111] structure is not only polar but even FE. But, polarization switching involves individual molecular rotation motions of 90$^{\circ}$ rather than a complete 180$^{\circ}$ switching of their orientation. 
Beyond this T-[111] configuration, alternating the possible $<$101$>$ orientations, we identified also various other polar and antipolar configurations at very close energies, highlighting the absence of strong correlation in the molecule orientations from one site to another. At room-temperature, the tetragonal phase is still observed to be disordered \cite{chen2015,leguy2015} and random orientation of the MA$^+$ molecules along equivalent $<$101$>$ directions preserves, on average, a non-polar character so that the system can still be seen as being in the paraelectric disordered regime. 

From this, MAPbI$_3$ should {\it a priori} exhibit on cooling an order-disorder transition toward the FE T-[111] phase at a given Curie temperature $T_{\text{FE}}$.  However, at $T$= 162K, appearance of additional PbI$_6$ octahedra rotations bring the system into a $Pnma$ orthorhombic phase. This phase is significantly lower in energy than the T-[111] phase and the combination of in-phase and out-of-phase AFD motions further distorts the A-site cavities in such a way that the MA$^+$ molecules are now forced to align along $<$100$>$ directions according to a well-defined antipolar pattern. The ground state of MAPbI$_3$ is therefore clearly not FE, although the rearrangement of MA$^+$ dipoles at ferroelastic domain walls might possibly contribute to provide some polar character.

In conclusion, MAPbI$_3$ can be seen as a missed FE in which the expected T-[111] FE tetragonal ground state is finally avoided by the appearance of AFD motions enforcing instead an antipolar orthorhombic ground state. The latter shows frozen orientations of dipoles, as confirmed experimentally \cite{weller2015}. Depending on how large is really $T_{\text{FE}}$, a FE T-[111] phase could eventually be stabilized in a narrow range of temperature above 162K. As previously mentioned, the molecules still rotate at room temperature \cite{chen2015,leguy2015}, locating $T_{\text{FE}}$ definitively below it. It has been reported that molecules' rotation significantly slows down at 190K \cite{koda2022} but no specific fingerprint of a FE phase transition has been observed at that temperature and molecules do not appear totally frozen yet. It could possibly happen that when the system finally reaches a temperature at which a more static polar configuration could  emerge, this helps initiating the final transition to the orthorhombic ground state that suppresses the FE transition. Although it does not seem to ever stabilize a FE state, we like to emphasize that, even in the paraelectric regime, MAPbI$_3$ can likely show fluctuating polar nanoregions -- which can become even more present below 190K -- and that poling it in a field might eventually bring it into a metastable FE state. This view can explain the controversial experimental observations on its FE character and contribute to clarify its puzzling properties. We mention finally that these conclusions reveal specific to MAPbI$_3$ and should not necessarily be seen as a generic property of similar hybrid perovskites. Nevertheless, the method and strategy we used here are versatile and might reveal appropriate to address the potential FE nature of related hybrid perovskites with other A-site molecules.

\section*{Methods}
\noindent
\textbf{\textit{Ab initio} calculations.} The calculations are performed within DFT using the full-potential projector augmented wave (PAW) method~\cite{blochl1994}, as implemented in the Vienna \textit{ab initio} Simulation Package (VASP)~\cite{kresse1996}. The exchange-correlation energy is estimated within the PBEsol \cite{perdew2008} generalized gradient approximation. For the tetragonal phase, we do further checks using the non-empirical strongly constrained and appropriately normed (SCAN) \cite{sun2015} meta-generalized gradient approximation (meta-GGA), which was shown to touch the accuracy of hybrid functional for diversely bonded molecules \cite{sun2016}. The relaxed states with different net polarizations are robust and indeed energetically comparable to PBEsol. We use a kinetic-energy cutoff of 500 eV. The \textit{k}-point mesh is adapted to the supercell under investigation. For the reference $1\times1\times1$ cubic structure, a $8\times8\times8$ $\Gamma$-centered Monkhorst-Pack  grid is adopted for the geometry optimization and self-consistent calculations. For the tetragonal and orthormbic phases, $\sqrt{2}\times\sqrt{2}\times2$ and $\sqrt{2}\times2\times\sqrt{2}$ supercells are used with $6\times6\times4$ and $6\times4\times6$ grids, respectively. The convergence criterion for the electronic energy is $10^{-6}$ eV and the structures are relaxed until the Hellmann–Feynman forces on each atom are less than 1 meV/\AA. The Berry-phase method \cite{king1993} is employed for the evaluation of the macroscopic polarization, properly eliminating the quantum indeterminacy \cite{vanderbilt1993}.

\bibliography{ref.bib}

\subsection*{\large{Acknowledgments}}
\noindent
W.-Y.T. acknowledges the support from F.R.S.-FNRS Belgium. J.-Z.Z acknowledges the support from the Startup Funding for Outstanding Young Scientist of South China Normal University and the financial support of China Scholarship Council (Grant No. 202006755025). The authors acknowledge access to the CECI supercomputer facilities funded by the F.R.S-FNRS (Grant No. 2.5020.1) and to the Tier-1 supercomputer of the Federation Wallonie-Bruxelles funded by the Walloon Region (Grant No. 1117545).

\subsection*{\large{Author contributions}}
\noindent
P.G. and J.-Z.Z. conceived the idea and supervised the work. W.-Y.T carried out first-principles calculations and did the data analysis. W.-Y.T. and J.-Z.Z. contributed to the interpretation of the results. W.Y.T and P.G. co-wrote the paper. All the authors reviewed and modified the manuscript.

\end{document}

\title{Supplementary Material - Missed ferroelectricity in methylammonium lead iodide}% Force line breaks with \\

\author{Wen-Yi Tong}
\affiliation{Theoretical Materials Physics, Q-MAT, CESAM, Universit\'e de Li\`ege, B-4000 Li\`ege, Belgium}
\author{Jin-Zhu Zhao}
\affiliation{Guangdong Provincial Key Laboratory of Quantum Engineering and Quantum Materials,Guangdong-Hong Kong Joint Laboratory of Quantum Matter,  School of Physics and Telecommunication Engineering,South China Normal University, Guangzhou 510006, P. R. China}
\affiliation{Center for Computational Science and Engineering, Southern University of Science and Technology, Shenzhen 518055, P. R. China}
\author{Philippe Ghosez}
\affiliation{Theoretical Materials Physics, Q-MAT, CESAM, Universit\'e de Li\`ege, B-4000 Li\`ege, Belgium}

\maketitle

\subsection{Polarization Details of Tetragonal Phases}

\begin{center}
\begin{tabularx}{\textwidth} { 
  >{\centering\arraybackslash}X| >{\centering\arraybackslash}X >{\centering\arraybackslash}X
  >{\centering\arraybackslash}X
  | >{\centering\arraybackslash}X
  >{\centering\arraybackslash}X
  >{\centering\arraybackslash}X
  | >{\raggedleft\arraybackslash}X
  >{\centering\arraybackslash}X
  >{\centering\arraybackslash}X
  | >{\centering\arraybackslash}X 
  >{\centering\arraybackslash}X
  >{\centering\arraybackslash}X }
  \hline
    \hline
       \multirow{2}*{} & \multicolumn{3}{c|}{\textbf{T-[111]}} & \multicolumn{3}{c|}{\textbf{T-[101]}} & \multicolumn{3}{c|}{\textbf{T-[001]}} & \multicolumn{3}{c}{\textbf{T-[100]}}\\
       \cline{2-13}
       &${P}_{\rm tot}$ &${P}_{\rm o-d}$ &${P}_{\rm dis}$ &${P}_{\rm tot}$ &${P}_{\rm o-d}$ &${P}_{\rm dis}$ &${P}_{\rm tot}$ &${P}_{\rm o-d}$ &${P}_{\rm dis}$ &${P}_{\rm tot}$ &${P}_{\rm o-d}$ &${P}_{\rm dis}$ \\
            \hline
     \textbf{$a$}  & 4.19 & 2.16 & 2.04 & 4.26 & 2.17 & 2.09 & 0.00 & 0.00 & 0.00 & 7.99 &  4.15 & 3.84  \\
     \textbf{$b$} & 4.19 & 2.16 & 2.04 &0.00 & 0.00 & 0.00 & 0.00 & 0.00 & 0.00 &  0.00 & 0.00 & 0.00  \\
     \textbf{$c$}  & -4.97 & -2.26 & -2.71 & -5.02 & -2.29 & -2.73 & -4.97 & -2.33 & -2.64 & 0.00 & 0.00 & 0.00 \\
     \hline
     module & 7.74 & 3.80 & 3.95 & 6.59 & 3.16 & 3.44 & 4.97 & 2.33 & 2.64 & 7.99 & 4.15 & 3.84\\
  \hline
 \hline
\end{tabularx}
\end{center}
TABLE SI. For the tetragonal states within different net polarization, components along $a$, $b$, and $c$ axis and the module of the total polarization ${P}_{\rm tot}$, and those of the ones related to the order-disorder ${P}_{\rm o-d}$ and the displacive ${P}_{\rm dis}$ mechanisms are reported in the unit of $\mu${C}/cm$^2$.

\subsection{Third- and Higher-order \textit{F} Terms}
                                 
\begin{center}
(a) 

      \begin{tabular}{ccc|c}
        \hline
        \hline
          \it{i} & \it{j} & \it{k} & $F_{ijk}$ \\
        \hline 
          $\varepsilon$ & $R$ & M$^\mathrm{o}_\mathrm{dis}$ & 1.26 \\
          $\varepsilon$ & $R$ & M$^\mathrm{o}_\mathrm{rot}$ & 5.40 \\
          $\varepsilon$ & $R$ & M$^\mathrm{i}_\mathrm{dis}$ & 7.10 \\
          $\varepsilon$ & $R$ & M$^\mathrm{i}_\mathrm{rot}$ & 6.57 \\
          $\varepsilon$ & M$^\mathrm{o}_\mathrm{dis}$ & M$^\mathrm{o}_\mathrm{rot}$ & -4.92 \\
          $\varepsilon$ & M$^\mathrm{o}_\mathrm{dis}$ & M$^\mathrm{i}_\mathrm{dis}$ & -6.15 \\
          $\varepsilon$ & M$^\mathrm{o}_\mathrm{dis}$ & M$^\mathrm{i}_\mathrm{rot}$ & -18.47 \\
          $\varepsilon$ & M$^\mathrm{o}_\mathrm{rot}$ & M$^\mathrm{i}_\mathrm{dis}$ & 5.40 \\
          $\varepsilon$ & M$^\mathrm{o}_\mathrm{rot}$ & M$^\mathrm{i}_\mathrm{rot}$ & -20.42 \\
          $\varepsilon$ & M$^\mathrm{i}_\mathrm{dis}$ & M$^\mathrm{i}_\mathrm{rot}$ & -26.65\\
        \hline
        \hline
      \end{tabular}
      \hspace{0.3in}
      \begin{tabular}{ccc|c}
        \hline
        \hline
          \it{i} & \it{j} & \it{k} & $F_{ijk}$ \\
        \hline 
          $R$ & M$^\mathrm{o}_\mathrm{dis}$ & M$^\mathrm{o}_\mathrm{rot}$ & 10.40 \\
          $R$ & M$^\mathrm{o}_\mathrm{dis}$ & M$^\mathrm{i}_\mathrm{dis}$ & -16.41 \\
          $R$ & M$^\mathrm{o}_\mathrm{dis}$ & M$^\mathrm{i}_\mathrm{rot}$ & -12.85 \\
          $R$ & M$^\mathrm{o}_\mathrm{rot}$ & M$^\mathrm{i}_\mathrm{dis}$ & 26.83 \\
          $R$ & M$^\mathrm{o}_\mathrm{rot}$ & M$^\mathrm{i}_\mathrm{rot}$ & -6.14 \\
          $R$ & M$^\mathrm{i}_\mathrm{dis}$ & M$^\mathrm{i}_\mathrm{rot}$ & 38.18\\
          M$^\mathrm{o}_\mathrm{dis}$ & M$^\mathrm{o}_\mathrm{rot}$ & M$^\mathrm{i}_\mathrm{dis}$ & 13.80\cellcolor{lightgray} \\
          M$^\mathrm{o}_\mathrm{dis}$ & M$^\mathrm{o}_\mathrm{rot}$ & M$^\mathrm{i}_\mathrm{rot}$ & 24.41\cellcolor{lightgray} \\
          M$^\mathrm{o}_\mathrm{dis}$ & M$^\mathrm{i}_\mathrm{dis}$ & M$^\mathrm{i}_\mathrm{rot}$ & -16.47\cellcolor{lightgray} \\
          M$^\mathrm{o}_\mathrm{rot}$ & M$^\mathrm{i}_\mathrm{dis}$ & M$^\mathrm{i}_\mathrm{rot}$ & 
          -2.74\cellcolor{lightgray} \\
        \hline
        \hline
      \end{tabular}
\end{center}

\begin{center}
(b)

      \begin{tabular}{cccc|c}
        \hline
        \hline
          \it{i} & \it{j} & \it{k} & \it{l} & $F_{ijkl}$ \\
        \hline 
          $R$ & M$^\mathrm{o}_\mathrm{dis}$ & M$^\mathrm{o}_\mathrm{rot}$ & M$^\mathrm{i}_\mathrm{rot}$ & -77.19 \\
          $R$ & M$^\mathrm{o}_\mathrm{dis}$ & M$^\mathrm{i}_\mathrm{dis}$ & M$^\mathrm{i}_\mathrm{rot}$ & -34.21 \\
          $R$ & M$^\mathrm{o}_\mathrm{rot}$ & M$^\mathrm{i}_\mathrm{dis}$ & M$^\mathrm{i}_\mathrm{rot}$ & 11.44 \\
          \hline
          $\varepsilon$ & M$^\mathrm{o}_\mathrm{dis}$ & M$^\mathrm{o}_\mathrm{rot}$ & M$^\mathrm{i}_\mathrm{rot}$ & 20.37 \\
          $\varepsilon$ & M$^\mathrm{o}_\mathrm{dis}$ & M$^\mathrm{i}_\mathrm{dis}$ & M$^\mathrm{i}_\mathrm{rot}$ & 8.91 \\
          $\varepsilon$ & M$^\mathrm{o}_\mathrm{rot}$ & M$^\mathrm{i}_\mathrm{dis}$ & M$^\mathrm{i}_\mathrm{rot}$ & 8.98 \\
        \hline
        \hline
      \end{tabular}
\end{center}
TABLE SII. (a) The third-order $F_{ij}$ and (b) parts of the forth-order $F_{ijkl}$ terms are listed here in the unit of meV/f.u.. Light-gray-highlighted data are in the cubic regime.